\documentclass[aps,prl,twocolumn,nofootinbib]{revtex4}
\usepackage{epsfig}
\usepackage{slashed}

\newcommand{\bel}[1]{\begin{equation}\label{#1}}
\newcommand{\bal}[1]{\begin{eqnarray}\label{#1}}

\newcommand{\be}{\begin{equation}}
\newcommand{\ee}{\end{equation}}
\newcommand{\ba}{\begin{eqnarray}}
\newcommand{\ea}{\end{eqnarray}}

\newcommand{\bes}{\begin{equation*}}
\newcommand{\ees}{\end{equation*}}

\begin{document}
\title{Finite-size scaling as a tool in the search for the QCD critical point in heavy ion data}
\author{Eduardo S. Fraga$^{1}$\footnote{fraga@if.ufrj.br},
Let\'\i cia F. Palhares$^{1,2}$\footnote{leticia@if.ufrj.br} and 
Paul Sorensen$^{3}$\footnote{psoren@bnl.gov}}
\affiliation{$^{1}$Instituto de F\'\i sica, Universidade Federal do Rio de Janeiro, \\
Caixa Postal 68528, Rio de Janeiro, RJ 21941-972, Brazil \\
$^{2}$Institut de Physique Th\'eorique, CEA/DSM/Saclay, Orme des Merisiers, 
91191 Gif-sur-Yvette cedex, France \\
$^{3}$Physics Department, Brookhaven National Laboratory, 
Upton, NY 11973-5000, USA}
\begin{abstract}
Given the short lifetime and the reduced volume of the quark-gluon plasma (QGP) formed in 
high-energy heavy ion collisions, a possible critical endpoint (CEP) will be blurred in a region 
and the effects from criticality severely smoothened. Nevertheless, the non-monotonic behavior 
of correlation functions near criticality for systems of different sizes, given by different centralities 
in heavy ion collisions, must obey finite-size scaling. We apply the predicting power of scaling 
plots to the search for the CEP of strong interactions in heavy ion collisions using data from RHIC 
and SPS. The results of our data analysis exclude a critical point below chemical potentials 
$\mu\sim 450~$MeV. Extrapolating the analysis, we speculate that criticality could appear slightly 
above $\mu\sim 500~$MeV. Using available data we extrapolate our scaling curves to predict the 
behavior of new data at lower center-of-mass energy, currently being investigated in the Beam 
Energy Scan program at RHIC. If it turns out that the QGP phase is no longer achievable in heavy 
ion experiments before the CEP is reached, FSS might be the only way to experimentally estimate 
its position in the phase diagram. 
\end{abstract}
\maketitle


High-energy heavy ion collisions can produce energy densities that are high enough 
to generate a new state of matter, the quark-gluon plasma (QGP), probing a new region in the 
phase diagram of strong interactions and testing quantum chromodynamics (QCD) under 
very extreme conditions. These experimental events are often referred to as 
{\it little bangs} since they reproduce some of the conditions found in the early universe at the 
time of the primordial quark-hadron transition. Nevertheless, one should bear in mind that 
the space-time scales that set the dynamics of the quark-gluon plasma formed in heavy ion 
collisions differ from the cosmological ones by almost twenty orders of magnitude. This means 
that the early universe at the time of the QCD phase transition(s) was already very large, in the 
sense that a description assuming the thermodynamic limit is well justified, whereas in the case 
of heavy ion collisions one should be more cautious. 

In fact, even if this issue is usually overlooked, it has been shown using lattice 
simulations \cite{Gopie:1998qn,Bazavov:2007zz} and different effective model 
approaches \cite{finite-NJL,Braun:2004yk,Yamamoto:2009ey,Palhares:2009tf} 
that finite-size effects can dramatically modify the phase structure of strong interaction, dislocating 
critical lines and critical points, and also affect significantly the dynamics of phase 
conversion \cite{Spieles:1997ab,Fraga:2003mu}. Moreover, as discussed in Ref. \cite{Palhares:2009tf}, 
the modifications suffered by the phase diagram features, such as the critical line and isentropic 
trajectories, are sensitive to the boundary conditions, as expected for small systems. Therefore, the 
data corresponding to small QGP systems, such as the ones formed in non-central heavy ion 
collisions, should not be naively compared  to what one would expect in the thermodynamic limit. 
A direct consequence of this fact is that all signatures of the second-order critical endpoint based 
on the non-monotonic behavior \cite{Stephanov:1998dy,Stephanov:2008qz} or sign modifications 
\cite{Asakawa:2009aj} of particle correlation fluctuations will probe a \textit{pseudocritical endpoint} 
that can be significantly shifted from the genuine (unique) critical endpoint by finite-size corrections 
and will be sensitive to boundary effects. This feature, together with the even more crucial limitation 
on the growth of the correlation length due to the finite (short) lifetime of the plasma state and 
critical slowing down \cite{Berdnikov:1999ph,Stephanov:2009ra}, makes the experimental searches 
of signatures of the presence of a critical point at lower energies very challenging, even if one 
considers higher moments which grow faster with the correlation length as suggested in 
Ref. \cite{Stephanov:2008qz}. The rounding and smoothening of fluctuation peaks tend to hide 
them behind the background. Besides, measuring higher moments of fluctuations of experimental 
observables in heavy-ion collisions is also a challenging task. 

On the other hand, given the fact that the thermal environment corresponding to the region of QGP 
formed in heavy ion collisions can be classified according to the centrality of the collision, so that 
events can be separated according to the size of the plasma that is created, the finiteness of the 
system also brings a bright side: the possibility of finite-size scaling (FSS) 
analysis \cite{fisher,Brezin:1981gm,Brezin:1985xx} (For reviews, see 
Refs. \cite{DL,Cardy:1996xt,amit}). FSS is a powerful statistical mechanics 
technique that prescinds from the knowledge of the details of a given system; instead, it provides 
information about its criticality based solely on very general characteristics. Although it is clearly 
not simple to define an appropriate scaling variable in the case of heavy ion collisions, the flexibility 
of the FSS method allows for a pragmatic approach for the use of scaling plots in the search for the 
critical endpoint as was delineated in Ref. \cite{Palhares:2009tf}. The essential point is that 
although the reduced volume of the QGP formed in high-energy heavy ion collisions will dissolve a 
possible critical point into a region and make the effects from criticality severely smoothened, the 
non-monotonic behavior of correlation functions for systems of different sizes, given by different 
centralities, must obey FSS near criticality \cite{DL,Cardy:1996xt,amit}. 

In this paper we apply the predicting power of scaling plots to the search for the CEP of 
strong interactions in heavy ion collisions using data from RHIC and SPS. Defining 
appropriate scaling variables, we generate scaling plots for data sets with 
$\sqrt{s_{NN}}=17.3, 19.6, 62.4, 130, 200$ GeV.
From the results of our data analysis we can exclude a critical point below chemical potentials 
$\mu\sim 450~$MeV. This seems to be consistent with lattice results that exclude a chiral critical 
point in the region of $\mu$ smaller or comparable to the temperature \cite{deForcrand:2007rq}. 

Extrapolating the analysis, we speculate that criticality could appear slightly above 
$\mu\sim 500~$MeV. Using available data we extrapolate our scaling curves to predict the 
behavior of new data at lower center-of-mass energy, currently being investigated in the Beam 
Energy Scan program at RHIC. If it turns out that the QGP phase is no longer achievable in heavy 
ion experiments before the CEP is reached, FSS might be the only way to experimentally estimate 
its position in the phase diagram. 

Finding the appropriate scaling variable(s) is almost never an easy exercise for real physical 
systems. In this analysis we pragmatically choose to work with a one-dimensional scaling 
function motivated by the simplification of thermal model descriptions of the freeze-out 
region \cite{Cleymans:2005xv}, which connects temperature and chemical potential, so that our 
scaling variable can be built from $\sqrt{s_{NN}}$. 


It is well known that as one goes near criticality in a second-order phase transition 
the correlation functions of the system scale with some power of the correlation length. 
This feature is independent of the details of the system under consideration, and the 
power is dictated by which correlation function one considers and the universality class 
to which the system belongs \cite{DL,Cardy:1996xt}. Therefore, it is clear that observables 
that can be written in terms of the correlation functions will exhibit a non-monotonic behavior 
near a critical point, the effect being more dramatic as one goes up in moments of the 
fluctuations. This behavior provides a clean evidence of the presence of a critical 
point \cite{DL,Cardy:1996xt}. Thus, it could be used as a precise signature in the analysis of data 
from heavy ion collisions using event-by-event analysis, as proposed in 
Refs. \cite{Stephanov:1998dy, Stephanov:2008qz}, were it not the case that these systems are 
usually very small and come in different sizes. Therefore, 
the peaks that are typical from this divergent behavior in the thermodynamic limit will be 
smoothened by the effects of finite size and short lifetime of the systems under consideration in 
heavy ion collisions. Even higher moments \cite{Stephanov:2008qz}, which could provide 
stronger divergences (that grow with higher powers of the correlation length) are more difficult to 
analyze under these circumstances, from the experimental point of view. So, as an alternative 
in data analysis, besides looking for non-monotonicity in particle multiplicities, one should also 
test for the scaling that differently-sized systems must obey near criticality, i.e. finite-size scaling. 
This also corresponds to the correct way to extrapolate the results of measurements of the finite 
system to the thermodynamic limit.

The FSS hypothesis \cite{fisher} was conjectured before the development of the renormalization 
group (RG). However, it can be derived quite naturally by applying RG techniques to critical 
phenomena \cite{Brezin:1985xx,amit}. For a system with typical linear dimension $L$, the singular 
part of the free energy density scales as
\begin{equation}
f_{s}(\{g\},L^{-1})=\ell^{-d}f_{s}(\{g'\},\ell L^{-1}) \; ,
\label{eq-fs1}
\end{equation}
when the lengths are reduced by a factor $\ell$ in a given RG transformation. $\{g\}$ are couplings 
and $d$ is the space dimension. The fact that the system is finite is irrelevant for this implementation, 
since the RG transformations are local. Near a RG fixed point, one can write (\ref{eq-fs1}) in terms of 
the right eigenvectors of the linearized RG transformation:
\begin{equation}
f_{s}(t,r,...,L^{-1})=\ell^{-d}f_{s}(t\ell^{y_{t}},r\ell^{y_{r}}...,\ell L^{-1})\;,
\label{eq-fs2}
\end{equation}
so that one notices that $L^{-1}$ behaves like a relevant eigenvector with eigenvalue 
$\Lambda_{L}=\ell$, and $y_{L}=1$. Here, $t=(T-T_{c})/T_{c}$ is the reduced temperature, 
$T_{c}$ being the temperature associated with the critical point. The reduced temperature 
$t$ is a dimensionless measure of the distance to the critical point when no other external 
parameter is considered. In the presence of other external parameters, such as the baryonic 
chemical potential $\mu$, one should redefine this distance accordingly
to include the dependence on $r=(\mu-\mu_{c})/\mu_{c}$, which plays a role analogous to the magnetic field variable in a Ising system.
$T_{c}$ and $\mu_{c}$ are defined in the thermodynamic limit, and 
the true second-order phase transition occurs for vanishing $t$, $r$ and $L^{-1}$ (and other 
couplings reaching their value at the fixed point, $\{g=g^{*}\}$). 

For finite $L$, crossover effects become important. If the correlation length diverges as 
$\xi_{\infty}\sim t^{-\nu}$ at criticality, where $\nu$ is the corresponding 
critical exponent, in the case of $L^{-1} t^{-\nu} \gg 1$ the system is no longer governed by the 
critical fixed point and  $L$ limits the growth of the correlation length, rounding all 
singularities \cite{goldenfeld}. If $L$ is finite, $\xi$ is analytic in the limit $t \to 0$, etc and one can 
draw scaling plots of $L/\xi$ vs. a given coupling $g$ for different values of $L$ to find that all curves 
cross at $g=g^{*}$ in this limit, which is a way to determine $g^{*}$. The critical temperature can 
also be determined in this fashion, since the curves will also cross at $t=t_{c}$.

This scaling plot technique can be extended, taken to its full power for other quantities, such as 
correlation functions. An observable $X$ in a finite thermal system can be written, 
in the neighborhood of criticality, in the following form \cite{Brezin:1985xx}:
\begin{equation}
X(t,\{g\};\ell;L)=L^{\gamma_{x}/\nu} f(t L^{1/\nu}) \;,  \label{scaling}
\end{equation}
where $\gamma_{x}$ is the bulk (dimension) exponent of $X$ and $\{g\}$ dimensionless coupling 
constants. The function $f(y)$ is universal up to scale fixing, and the critical exponents are sensitive 
essentially to dimensionality and internal symmetry, which will give rise to the different universality 
classes \cite{DL,amit}. To simplify the discussion, we ignore for the moment the chemical potential 
variable $r$. In principle, $f$ is a function of two scaling variables, though. Again, if one plots the 
scaled observable versus $t$, the curves should cross at $t=t_{c}$. Moreover, using the appropriate 
scaling variable, instead of $t$, all curves should collapse into one single curve if one is not far 
from the critical point. So, this technique can be applied to the analysis of observables  that are 
directly related to the correlation function of the order parameter of the transition, such as fluctuations 
of the multiplicity of soft pions \cite{Stephanov:1998dy}. 

This is the idea of using full scaling plots, a method that has been proven to work even for tiny systems, 
to search for the critical endpoint of QCD in heavy ion data. Here, as in statistical mechanics searches 
for critical points, one can treat $T_{c}$, $\mu_{c}$ and the critical exponents as parameters in a scaling 
plot fit, the system size being provided by different centrality bins or by the number of participants in the collision. In fact, the scaling variable is defined up to $L$- and $t$-independent multiplicative factors, 
so that the knowledge of the actual size of the system is not needed.

As mentioned previously, the correct scaling variable should measure the distance from the 
critical point, thereby involving both temperature and chemical potential, or their reduced versions 
$t$ and $r$. This produces a two-dimensional scaling function and makes the analysis of heavy 
ion data highly nontrivial. Phenomenologically, we adopt a simplification motivated by results 
from thermal models for the freeze-out region, connecting temperature and chemical potential. 
We can parametrize the freeze-out curve by $\sqrt{s_{NN}}$, and build our one-dimensional 
scaling variable from this quantity and the size of the system. 


The range of sizes that can be accessed in heavy ion collisions over which FSS can be tested is limited. It is reasonable to assume that for a locally thermalized quark gluon plasma to form, the system needs to be several times the hadronic size (several fm). The largest possible system will have a diameter of approximately $15$ fm. Estimates from an analysis of HBT data from STAR indicate that from peripheral to central collisions, the system size changes by a factor of $3-4$~\cite{hbtsize}. We plot the scaled observable vs $\left[(\mu-\mu_c)/\mu_c\right]L^{1/\nu}$.  The value of $L$ for the various centralities can be estimated from a Glauber Monte-Carlo model. For our study, the overlap area is calculated from $S_{\perp}=R_{Au}^2(\Theta-\sin\Theta)$ where $\Theta=2\cos^{-1}(b/2R_{Au})$, where $R_{Au}$ is the nuclear radius for Au and $b$ is the collision impact parameter. Then the length is taken to be $L=2\sqrt{S_{\perp}/\pi}$.

A system of size $L_1$ can be compared to a system of size $L_2$ only when
\begin{equation}
  \mu_1-\mu_c=(\mu_2-\mu_c)\left( \frac{L_2}{L_1} \right)^{1/\nu}.
\end{equation}
This constrains which $\mu$ values can be compared when testing for scaling by directly comparing data rather than by extrapolations. Since a central $Au+Au$ collision is about four times larger than a peripheral one ($L_{2}=4L_{1}$) and taking $\nu=2/3$, we find that a measurement in peripheral Au+Au collisions at $\mu_1$ can be compared to a central collision at $\mu_2$ when $\mu_1-\mu_c\approx8(\mu_2-\mu_c)$.

Data on various fluctuations have been measured at RHIC and the SPS over an energy range 
from $\sqrt{s_{NN}}=5$ GeV to $200$ GeV and the centrality dependence of $p_T$ fluctuations 
have been measured at RHIC for energies from $19.6$ GeV to $200$ GeV. This data provides an 
opportunity to look for evidence of finite-size scaling. 

To search for scaling, we consider the correlation measure $\sigma_{p_T}/\langle p_T\rangle$~\cite{Adams:2005ka} 
scaled by $L^{-\gamma_x/\nu}$, according to Eq. \ref{scaling}. We consider the $p_T$ fluctuations 
$\sigma_{p_T}$ scaled by $\langle p_T\rangle$ to obtain a dimensionless variable \footnote{
The STAR collaboration has also published $p_T$ fluctuations using an alternative variable 
$\Delta\sigma_{p_T}$~\cite{Adams:2006sg}. The results found using this variable are qualitatively equivalent to the ones for 
$\sigma_{p_T}/\langle p_T\rangle$.}. We use the correlation data measured in bins corresponding to the 0-5\%, 5-10\%, 10-20\%, 20-30\%, 30-40\%, 40-50\%, 50-60\%, and 60-70\% most central collisions. We estimate the corresponding lengths L to be 12.4, 11.1, 9.6, 8.0, 6.8, 5.6, 4.5, and 3.4 fm.
%
%
The exponent $\nu = 2/3$ is determined by the Ising universality 
class of QCD and we consider values of $\gamma_{x}$ around $1$ (ignoring small anomalous 
dimensions corrections). 
We also varied the value of $\gamma_{x}$ from $0.5$ to $2.0$ and found that changing $\gamma_{x}$ 
within this range does not improve the scaling behavior. Using a smaller value of $\gamma_{x}$ 
simply causes $\sigma_{p_T}/\langle p_T\rangle$ scaled by $L^{-\gamma_x/\nu}$ to drop more 
slowly with increasing centrality and increasing $\gamma_x$ causes this drop to happen more 
quickly.

\begin{figure}[htb]
\centering\mbox{
\includegraphics[width=0.47\textwidth]{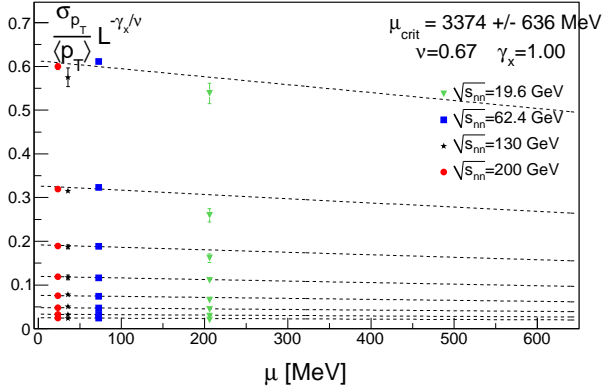}}
\caption[]{ Scaled $\sigma_{p_T}/\langle p_T\rangle$ vs $\mu$ for different system sizes, and 
with $\nu=2/3$ and $\gamma_{x}=1$. Data extracted from RHIC collisions at energies 
$\sqrt{s_{NN}}=19.6, 62.4, 130$, and $200$ GeV (linear fit, see text). }
\label{scaling-linear}
\end{figure}

\begin{figure}[htb]
\centering\mbox{
\includegraphics[width=0.47\textwidth]{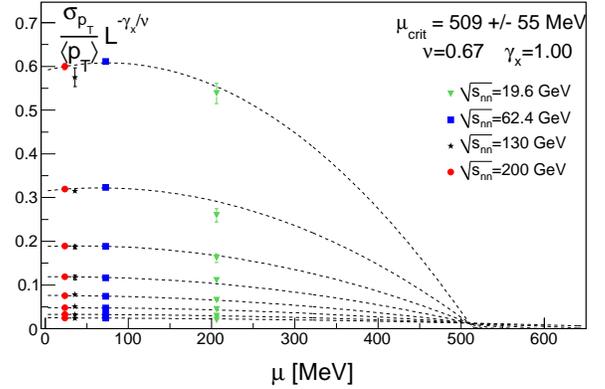}}
\caption[]{ Scaled $\sigma_{p_T}/\langle p_T\rangle$ vs $\mu$ 
for different system sizes, assuming $\mu_{\mathrm{crit}}=509$ MeV, which corresponds to a critical point at 
$\sqrt{s_{NN}}=5.75$ GeV. Again, $\nu=2/3$ and $\gamma_{x}=1$.
Data extracted from RHIC collisions at energies $\sqrt{s_{NN}}=19.6, 62.4, 130$, and $200$ GeV 
(second order polynomial fit, see text).}
\label{scaling-pol2}
\end{figure}

In Figs.~\ref{scaling-linear} and \ref{scaling-pol2} we plot $\sigma_{p_T}/\langle p_T\rangle$ scaled by 
$L^{-\gamma_x/\nu}$ vs $\mu$ for different system sizes, using data extracted from collisions at 
$\sqrt{s_{NN}}= 19.6, 62.4, 130, 200$ GeV. We use the parameterization from Ref.~\cite{conversion} to convert from $\sqrt{s_{NN}}$ to $\mu$. If there is a critical point at $\mu=\mu_{\mathrm{crit}}$, the 
curves for different sizes of the system should cross at this value of $\mu$. However, since the 
currently available data is restricted to not so large values of the chemical potential, one has to 
perform extrapolations using fits. The scaling function $f$ in Eq. \ref{scaling} is expected to be 
smoothly varying around the critical point, so we fit the data corresponding to a given linear size $L$ to 
a polynomial, but constraining the polynomials to enforce the condition that all the curves cross at 
some $\mu=\mu_{\mathrm{crit}}$, where $\mu_{\mathrm{crit}}$ is an adjustable parameter in the fit. This clearly assumes 
the existence of a critical point. In Fig.~\ref{scaling-linear} we use a linear fit. The approximate energy independence of $\sigma_{p_T}/\langle p_T\rangle$ along with the linear fit, leads to a very large $\mu$ value where the curves can cross ($\mu\sim 3~$GeV).  There's no reason however to assume a linear fit function, so in Fig.~\ref{scaling-pol2} we also try a second order polynomial. Using the a second order polynomial function for $f$ allows the curves from different system sizes to cross at a much smaller value of $\mu$. Based on this fit, we find that the data is consistent with a critical point at $\mu\sim510~$MeV corresponding to a $\sqrt{s_{NN}}$ of 5.75 GeV.

The value estimated for $\mu_{\mathrm{crit}}$ based on finite-size scaling of current data is highly dependent on the assumed functional form of $f$. The approximate energy independence of $\sigma_{p_T}/\langle p_T\rangle$ for a given L, however, already indicates within the finite-size scaling assumption that the critical point should be at $\mu$ values well above those currently available. Based on finite-size scaling of $\sigma_{p_T}/\langle p_T\rangle$, one would not expect a critical point at $\mu<400~$MeV. Having data for lower values of $\sqrt{s}$, i.e. higher values of chemical potential, as expected from the analysis of the Beam Energy Scan program at RHIC~\cite{starbes}, one should be able to study full scaling plots of $\sigma_{p_T}/\langle p_T\rangle$ scaled by $L^{-\gamma_x/\nu}$ vs. $\frac{\mu-\mu_{\mathrm{crit}}}{\mu_{\mathrm{crit}}}L^{1/\nu}$ without the need of long extrapolations. For the current set of data, these full scaling plots are still not very enlightening.

\begin{figure}[htb]
\centering\mbox{
\includegraphics[width=0.47\textwidth]{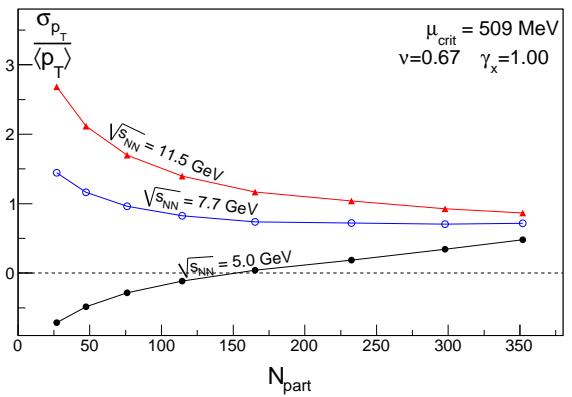}}
\caption[]{ The expected measurement of $\sigma_{p_T}/\langle p_T\rangle$ as a function 
of the number of participants at lower energies assuming the critical point is at $509$ MeV 
as extracted from the quadratic polynomial fit of STAR data.}
\label{predict}
\end{figure}

RHIC has also run at lower energies in order to search for a critical point in the Beam Energy 
Scan program.  That data is currently being analyzed. Here we use the quadratic polynomial fit 
of STAR data (Fig. \ref{scaling-pol2}) and assume the critical point is at $509$ MeV to make 
predictions for $\sigma_{p_T}/\langle p_T\rangle$ at lower energies. The finite-size scaling 
scenario along with currently available data, allows us to predict the energy and system size 
dependence of fluctuations for any given values of $\mu_{\mathrm{crit}}$, $\gamma_x$ and $\nu$.  
We show this expectation as a function of the number of participants, $N_{\mathrm{part}}$, for three 
proposed beam energies: $11.5, 7.7$ and $5$ GeV in Fig.~\ref{predict}. Notice that the 
centrality dependence changes once one moves to the other side of the critical point.
This is a condition enforced by finite-size scaling which provides a generic signal for having reached the first-order phase transition side of the critical point.

%


Searching for the the critical endpoint in the phase diagram for strong interactions is a remarkably 
challenging task. On one hand, lattice simulations can not provide the sort of guidance that is possible 
at zero density, where there is no sign problem, and effective models point to its existence, qualitatively, 
but yield very different quantitative predictions. On the other hand, heavy ion collision experiments 
probe limited regions of the phase diagram. Besides, they come with non-trivial background contributions 
that tend to blur the signatures provided by the non-monotonic behavior of observables built from 
correlation functions of the order parameter.

The fact that finite-size scaling prescinds from the knowledge of the details of the system under 
consideration, providing information about its criticality based solely on its most general features, 
makes it a very powerful and pragmatic tool for data analysis in the heavy ion collision search for 
the critical point. From a very limited data set in energy spam, we have used FSS to exclude the 
presence of a critical point at small values of the baryonic chemical potential, below $450~$MeV. 
We have also used the scaling function to predict the behavior of data with system size at lower 
energies. We are looking forward to compare our predictions to the outcome of data analysis from 
the Beam Energy Scan program at RHIC.

We are grateful to M. Chernodub, T. Kodama, \'A. M\'ocsy, K. Rajagopal, K. Redlich and M. Stephanov 
for fruitful discussions. This work was partially supported by CAPES-COFECUB (project 663/10), 
CNPq, FAPERJ and FUJB/UFRJ.


\end{document}